\documentclass{aa}

\usepackage{amsmath}
\usepackage{graphicx}
\usepackage[mathscr]{euscript} 
\usepackage{comment}
\usepackage[utf8]{inputenc}

\usepackage{booktabs}
\usepackage{adjustbox}
\usepackage{tablefootnote}

\usepackage{caption}
\usepackage{subcaption}
\usepackage{ulem}
\usepackage{dirtytalk}

\usepackage{multicol}
\usepackage{multirow}
\usepackage{txfonts}
\usepackage{natbib} 
\bibpunct{(}{)}{;}{a}{}{,}

\usepackage{breqn} 
\usepackage[colorlinks=true, linkcolor=blue, citecolor=blue, urlcolor=blue]{hyperref}

\begin{document}

   \title{Electromagnetic signatures of\\ black hole clusters in the center of super-Eddington galaxies}
   \author{Leandro Abaroa\inst{1,2}\thanks{leandroabaroa@gmail.com} \and  
          Gustavo E. Romero\inst{1,2}
           }

   \offprints{Leandro Abaroa}
  \institute{Instituto Argentino de Radioastronom\'ia (CCT La Plata, CONICET; CICPBA; UNLP), C.C.5, (1894) Villa Elisa, Argentina \and Facultad de Ciencias Astron\'omicas y Geof\'{\i}sicas, Universidad Nacional de La Plata, B1900FWA La Plata, Argentina}

   \date{Received / Accepted}

\abstract{
Supermassive black holes (SMBHs) at the centers of active galaxies are fed by accretion disks that radiate from the infrared or optical to the X-ray bands. Several types of objects can orbit SMBHs, including massive stars, neutron stars, clouds from the broad- and narrow-line regions, and X-ray binaries. Isolated black holes with a stellar origin (BHs of $\sim10\,M_{\odot}$) should also be present in large numbers within the central parsec of the galaxies. These BHs are expected to form a cluster around the SMBH as a result of the enhanced star formation rate in the inner galactic region and the BH migration caused by gravitational dynamical friction. However, except for occasional microlensing effects on background stars or gravitational waves from binary BH mergers, the presence of a BH population is hard to verify. In this paper, we explore the possibility of detecting electromagnetic signatures of a central cluster of BHs when the accretion rate onto the central SMBH is greater than the Eddington rate. In these supercritical systems, the accretion disk launches powerful winds that interact with the objects orbiting the SMBH. Isolated BHs can capture matter from this dense wind, leading to the formation of small accretion disks around them. If jets are produced in these single microquasars, they could be sites of particle acceleration to relativistic energies. These particles in turn are expected to cool by various radiative processes. Therefore, the wind of the SMBH might illuminate the BHs through the production of both thermal and nonthermal radiation. We conclude that, under these circumstances, a cluster of isolated BHs could be detected at X-rays (with \textit{Chandra} and \textit{XMM-Newton}) and radio wavelengths (e.g., with the Very Large Array and the Square Kilometer Array) in the center of nearby super-Eddington galaxies.}

\authorrunning{L. Abaroa \& G.E. Romero}
\titlerunning{Black holes in super-Eddington galactic centers}

\maketitle
\nolinenumbers


\section{Introduction}

Supermassive black holes (SMBHs) at the centers of active galaxies are fed by accretion disks that radiate from the optical band to the X-rays. In many cases, a dusty torus obscures the central region and emits mainly in the infrared. 

Various types of objects can orbit the SMBHs at the centers of galaxies, including massive stars \citep{1976ApJ...209..214B}, neutron stars \citep{2018JKAS...51..165K}, clouds from broad- and narrow-line regions \citep{2020A&A...636A..92M, 2022A&A...664A.178S}, and X-ray binaries \citep{Hailey_etal2018,2021ApJ...921..148M}. Isolated black holes with a stellar origin (BHs of $\sim 10 M_{\odot}$) should also be present in large numbers within the central parsec of most galaxies. These BHs are expected to form a cluster around the SMBH as a result of the enhanced star formation rate in the inner galaxy and BH migration caused by gravitational dynamical friction \citep{2000ApJ...545..847M,2011MNRAS.413L..24M,2019MNRAS.482.3669G}. However, except for occasional microlensing effects on background stars \citep{2001ApJ...563..793C} or gravitational waves from BH binary mergers \citep{2019ApJ...881...20R,2022MNRAS.515.3299G,gaete2024supermassive}, the presence of a BH population is hard to verify. Observational signatures associated with the accretion of matter onto the BHs may provide clues for identifying them.

Accretion of matter onto BHs on all scales proceeds in three basic regimes, depending on the relation of the actual accretion rate $(\dot{M}_{\rm input})$ to the Eddington rate $(\Dot{M}_{\mathrm{Edd}})$: a highly sub-Eddington regime, $\Dot{M}_{\rm input} \ll \Dot{M}_{\mathrm{Edd}}$ \citep{1994ApJ...428L..13N}; a moderate sub-Eddington regime, $\Dot{M}_{\rm input} \lesssim \Dot{M}_{\mathrm{Edd}}$ \citep{1973A&A....24..337S}; and a super-Eddington regime, $\Dot{M}_{\rm input} \gg \Dot{M}_{\mathrm{Edd}}$ \citep{1980ApJ...242..772A,2004PASJ...56..569F}. In the last regime, the radiation pressure overcomes gravity in the innermost part of the disk, which becomes geometrically and optically thick. The surface layers of the disk are ejected as a consequence of the self-regulation of the BH accretion at the Eddington rate \citep{2009PASJ...61.1305F,2015PASJ...67..111T,2023A&A...671A...9A}. Two examples of super-Eddington accretion systems involving stellar BHs are ultraluminous X-ray sources \citep[e.g. NGC4190 X-1, see][]{2023A&A...671A...9A, Combi_etal_2024} and some microquasars, such as the Galactic SS433 \citep{2004ASPRv..12....1F}. 
SMBHs can also accrete at supercritical rates in some cases \citep[e.g. Mrk 335,][]{2022A&A...664A.178S}. 

The actual fraction of supercritical SMBHs in the galaxies is not well known \citep[see][for some samples]{2016ApJ...825..126D,2018ApJ...856....6D,2021ApJ...910..103L}.  In these systems, the accretion disk surrounding the central SMBH launches powerful winds with mass-loss rates similar to the accretion rate. In particular, when the SMBH has a relatively low mass, $\lesssim 10^7 M_{\odot}$, it can disrupt stars by tidal forces (in a tidal disruption event; TDE) triggering supercritical \citep{2012ApJ...760..103D,2016MNRAS.455..859S,Tadhunter_etal_2017_nature,Zubovas_2019,2021SSRv..217...12D,kaur2024elevated,2024arXiv240409381P} or hypercritical accretion regimes with $\dot{M}_{\rm SMBH}\sim10^3-10^4\dot{M}_{\rm Edd}$ in the most extreme cases \citep{2016MNRAS.459.3738I,2016MNRAS.461.4496S,2020MNRAS.497..302T}. These strong outflows then interact with the 
objects orbiting the SMBH on their way out of the system.

Isolated BHs from the central cluster can capture matter from the dense radiation-driven wind, leading to the formation of small accretion disks around them via the Bondi-Hoyle-Lyttleton (BHL) mechanism \citep[for disk formation in BHL regimes, see e.g.][]{Kolykhalov&Syunyaev1979,Soker2004,Okazaki_etal2008,Romero-Vila2014,Mellah_etal2019,Hirai&Mandel2021}. The small accretion disks produce thermal radiation, and when jets are formed in these "single" microquasars, they can be sites of particle acceleration to relativistic energies \citep[see][for jet formation in the BHL regime]{2023ApJ...950...31K}. These particles in turn are expected to cool by synchrotron radiation and inverse-Compton (IC) upscattering of photons, emitting nonthermal radiation. Proton acceleration can also lead to gamma-ray emission through $pp$ collisions and subsequent $\pi^0$ decay 
\citep{2003A&A...410L...1R,2008A&A...485..623R,2019A&A...629A..76S}. Furthermore, the bow shock that forms at the head of the propagating jet can also be a source of optically thin thermal radiation.
Therefore, the wind of the SMBH might illuminate the stellar BHs through the production of both thermal and nonthermal radiation. This radiation thus becomes a tracer of the presence of the BH cluster. Fig. \ref{fig: sketch} shows a scheme of the proposed scenario. 

Accretion onto isolated stellar BHs has been studied by \cite{2018MNRAS.477..791T} and \cite{2019MNRAS.489.2038I}. In particular, \cite{Maccarone2005}, \cite{Fender_etal2013}, and \cite{2019MNRAS.488.2099T} showed that these BHs can be detected, and \cite{2012MNRAS.427..589B} demonstrated that jets can be produced despite the very low accretion rates involved. On the other hand, observable signatures of stellar mass BHs embedded in SMBH disks have recently been studied by \cite{Tagawa_etal2023b}, \cite{Tagawa_etal2023a}, and  \cite{2024arXiv240709945R}.

In this paper, we explore the possibility of detecting electromagnetic signatures from a cluster of BHs orbiting a super-Eddington SMBH without jets, as in the case of some Seyfert 1 galaxies. We consider an SMBH in the super-Eddington regime (either a persistent regime or a transient regime generated by a TDE) that ejects winds, and we characterize the accretion of the dense wind onto the stellar BHs and the subsequent formation of jets.  We calculate the thermal emission from the disk, the nonthermal emission from the inner jet, and the thermal and nonthermal radiation from the terminal regions of the jet. We then correct the emission for absorption locally and in the wind photosphere. Finally, we consider whether the integrated spectral energy distribution (SED) shows any features that could reveal the presence of a BH cluster at the center of some galaxies.

The paper is organized as follows. We begin by characterizing the astrophysical scenario in which stellar BHs orbit the SMBH, decribing its disk and wind, the population of BHs, and the accretion of an isolated BH moving in a dense wind (Sect. \ref{sec: model}). In Sect. \ref{sec: jet} we study the production of jets in BHs and their interaction with the SMBH wind. We then describe the associated radiative processes and the absorption of the radiation. The results are presented in Sect. \ref{sec: results}, and after a discussion (Sect. \ref{sec: discussion}), we conclude.

\section{Model} \label{sec: model}

\begin{figure*}
    \centering
    \includegraphics[width=18.3cm]{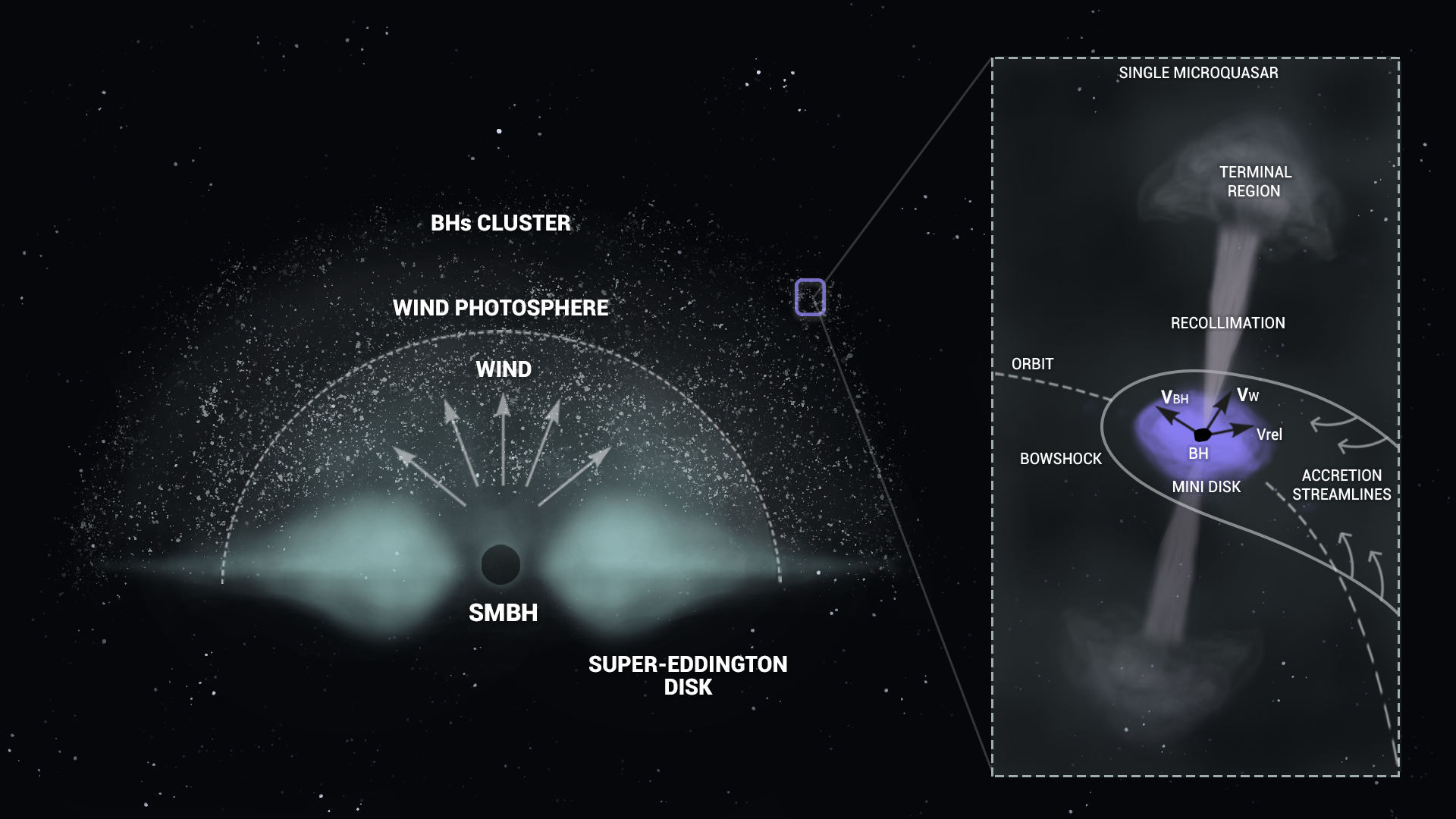}
    \caption{Scheme of the proposed scenario (not to scale). The wind ejected from the accretion disk of the SMBH interacts with the BHs forming the cluster. Because of the combination of the velocities of the SMBH wind $(v_{\rm w})$ and the orbital motion of the BH $(v_{\rm BH})$, a bow shock forms in the direction of the relative velocity $(v_{\rm rel})$. The dense wind is captured by the BH via cylindrical accretion. A small accretion disk forms around the BH and jets are produced with two assumed shock-prone zones (recollimation and terminal regions).}
    \label{fig: sketch}
\end{figure*}

We assumed that the BHs from the cluster have typical masses of $M_{\rm BH}\sim10 M_{\odot}$ \citep{2000ApJ...545..847M,2011MNRAS.413L..24M} and orbit a superaccreting nonjetted SMBH of $M_{\rm SMBH}=10^7 M_{\odot}$ in circular and Keplerian orbits. The stellar BHs accrete matter from the dense wind ejected from the accretion disk of the SMBH. The accretion onto the BHs proceeds via the BHL regime, and jets are generated by magneto-centrifugal effects \citep[e.g.,][]{2005ApJ...629..960S}. Fig. \ref{fig: sketch} shows a scheme of this scenario. In what follows, we first describe the SMBH and its wind, then the cluster of BHs orbiting the central SMBH, and finally, the accretion onto the BHs, considering three different scenarios. Table \ref{tab: system} summarizes the values of the main parameters of the system.

\begin{table}
\caption{Basic parameters of the SMBH, BHs, and cluster.}
\begin{center}
\begin{adjustbox}{width=\columnwidth}
\begin{tabular}{l c c}
\hline
\hline
\rule{0pt}{2.5ex}Parameter & Symbol\,[units] & Value \\
\hline
\rule{0pt}{2.5ex}Supermassive black hole mass$^{(1)}$ & $M_{\rm SMBH}$ [$M_{\odot}$]  & $10^7$  \\
Gravitational radius (SMBH)$^{(2)}$ & $r_{\rm g,SMBH}$ [$\rm{cm}$] & $1.48\times10^{12}$ \\
Eddington rate (SMBH)$^{(2)}$ & $\dot{M}_{\rm Edd,SMBH}$  [$\rm g\,{s}^{-1}$] & $1.4\times10^{25}$  \\
Cluster radius$^{(2)}$ & $r_{\rm cluster}$  [$\rm{cm}$] & $1.6\times10^{19}$ \\
Population of BHs in the cluster$^{(2)}$ & $N_{\rm BH}$ & $1.5\times 10^5$ \\
Stellar black hole mass$^{(1)}$ & $M_{\rm BH}$  [$M_{\odot}$]  & $10$ \\
Gravitational radius (BH)$^{(2)}$ & $r_{\rm g,BH}$  [$\rm{cm}$] & $1.48\times10^6$ \\
Eddington rate (BH)$^{(2)}$ & $\dot{M}_{\rm Edd,BH}$  [$\rm g\,{s}^{-1}$] & $1.4\times10^{19}$ \\
\hline
\end{tabular}
\end{adjustbox}
\end{center}
\tablefoot{We indicate with superscript ${(1)}$ the parameters that were assumed and with ${(2)}$ those that were derived.} \label{tab: system}
\end{table}

\begin{table*}
\begin{center}
\caption{Parameters of the three scenarios we modeled for the accretion rate onto the SMBH.}
\label{tab: scenarios}
\begin{tabular}{p{6cm} p{3cm} c c c }
\hline
\hline
\rule{0pt}{2.5ex}Parameter & Symbol\,[units] & $\zeta_1$ & $\zeta_2$ & $\zeta_3$   \\
\hline
\rule{0pt}{2.5ex}Mass accretion rate of the SMBH & $\dot{M}_{\rm input}$  [$\dot{M}_{\rm Edd,SMBH}$] & $10$ & $10^{2}$ & $10^{3}$ \\
Mass loss rate of the SMBH disk in winds & $\dot{M}_{\rm w}$  [ ${\rm g\,{s}^{-1}}$] & $1.4\times10^{26}$ & $1.4\times10^{27}$ & $1.4\times10^{28}$  \\
Wind velocity & $v_{\rm w}$  [${\rm cm\,{s}^{-1}}$] & $1.5\times10^{9}$ & $4.8\times10^{8}$ & $1.5\times10^{8}$  \\
Height of the wind photosphere & $z_{\rm photo}$  [$r_{\rm g,SMBH}$] & $1.58\times10^{2}$ & $6.3\times10^{3}$ & $1.58\times10^{5}$   \\
Temperature of the wind photosphere & $T_{\rm photo}$  [${\rm K}$] & $6.3\times10^{4}$ & $1.58\times10^{4}$ & $2\times10^{3}$ \\
\hline
\end{tabular}
\tablefoot{$\dot{M}_{\rm input}$ was assumed and the rest of the parameters were derived.} 
\end{center}
\end{table*}

\subsection{Accretion disk of the SMBH and its wind} \label{sec: SMBH disk and wind model}

We neglected the self-gravity of the disk of the SMBH, which is under a super-Eddington accretion regime in the outer part of the disk, $\zeta=\dot{M}_{\rm input}/\dot{M}_{\rm Edd,SMBH} > 1$, where $\dot{M}_{\rm input}$ is the supply of mass per unit time. We explored three super-Eddington regimes with rates of $\zeta=10,10^2$, and $10^3$. The Eddington rate is given by
\begin{equation} \label{eq: tasa eddington}
    \Dot{M}_{\rm{Edd}}= \frac{L_{\rm{Edd}}}{\eta c^2} \approx 2.2\times 10^{-8} M \, {\rm yr^{-1}} = 1.4 \times 10^{18} \frac{M}{M_\odot} \ \rm{g \, s^{-1}},
\end{equation}
where $L_{\rm Edd}$ is the Eddington luminosity (defined as the luminosity required to balance the gravitational attraction of the accreting object by radiation pressure), $\eta \approx 0.1$ is the accretion efficiency, $M$ is the accretor mass in units of solar mass, and $c$ is the speed of light. 

In the cases of high accretion rates $(\zeta{\gtrsim} 10^2)$, we assumed that the super-Eddington regime of the SMBH is caused by the disruption of a star in a TDE. During a TDE, the peak mass fallback rate is given by \citep{2016MNRAS.455..859S}
 \begin{equation}
     \dot{M}_{\rm peak}=4.2 \left(\frac{\eta}{0.1}\right)\left(\frac{M_{\rm SMBH}}{10^7\,M_{\odot}}\right)^{-3/2}\left(\frac{M_{*}}{M_{\odot}}\right)^2 \left(\frac{R_{*}}{R_{\odot}}\right)^{-3/2} \ \dot{M}_{\rm Edd},
 \end{equation}
with $M_*$ the mass and $R_*$ the radius of the disrupted star. This super-Eddington phase remains for a timescale
\begin{equation} \label{eq: t edd}
    t_{\rm Edd}= 0.83 \left(\frac{\eta}{0.1}\right)^{3/5}\left(\frac{M_{\rm SMBH}}{10^7\,M_{\odot}}\right)^{-2/5}\left(\frac{M_{*}}{M_{\odot}}\right)^{1/5} \left(\frac{R_{*}}{R_{\odot}}\right)^{3/5} {\rm yr}.
\end{equation}

The central parsecs of galaxies are expected to harbor early massive stars, which means that when we calculate the accretion peak for an O5 star with mass $37\,M_{\odot}$ and radius $11\,R_{\odot}$ \citep{2019AJ....158...73K}, for example, we obtain $\dot{M}_{\rm peak}>10^2 \dot{M}_{\rm Edd}$, a superaccretion regime that lasts for a time $t_{\rm Edd}\approx 7\ {\rm yr}$. 

The critical radius of the disk, given by $r_{\rm crit} \sim  40 \,\zeta \,r_{\rm g, SMBH}$, corresponds to the distance from the SMBH at which the outer standard disk \citep{1973A&A....24..337S} changes to the radiation-dominated inner disk \citep{2004PASJ...56..569F}. Here, $r_{\rm g,SMBH}=GM_{\rm SMBH}/c^2$ is the gravitational radius of the SMBH, with $G$ the gravitational constant. 
The disk becomes geometrically thick in the inner region, where the ejection of winds by the radiation force helps to regulate the mass accretion rate onto the SMBH ($\dot{M}_{\rm acc}$) at the Eddington rate.\footnote{$\dot{M}_{\rm acc}=\dot{M}_{\rm input}$ in the outer region of the disk and $\dot{M}_{\rm acc}=\dot{M}_{\rm input}r_{\rm d}/r_{\rm crit}$ in the inner region, where $r_{\rm d}$ is the distance to the SMBH in the equatorial plane of the disk \citep{2004PASJ...56..569F}.} This regulation results in a total mass-loss rate $\dot{M}_{\rm w}$ that is approximately equal to the accretion input, $\dot{M}_{\rm w}\approx \dot{M}_{\rm input}=\zeta\dot{M}_{\rm Edd,SMBH}$. 




We assumed that the wind ejected from the SMBH disk is smooth and expands isotropically at a constant velocity, estimated to be $\beta_{\rm w}=v_{\rm w}/c=1/\sqrt{40\, \dot{m}}$ \citep{2010MNRAS.402.1516K}. Since $\beta_{\rm w}\ll1$ for $\zeta\gg1$, the Lorentz factor of the wind is $\gamma_{\rm w}\approx1$, and therefore, the wind properties do not depend on the line of sight ($\Theta$), so that $1-\beta_{\rm w} \cos \Theta \approx1$ \citep[see e.g.,][]{Abaroa&Romero_2024RevMex}. The density of this wind at a distance $r$ from the BH is
\begin{equation} \label{eq: densidad_viento}
    \rho_{\rm w}(r)=\dot{M}_{\rm w}/4\pi r^2 v_{\rm w}.
\end{equation}
We assumed steady state, so that $v_{\rm w}$ and $\dot{M}_{\rm w}$ are constant, and therefore, the density only depends on the distance to the SMBH.

The wind itself is opaque until it reaches the photospheric radius, where it becomes transparent to its own radiation. The apparent photosphere is defined as the surface where the optical depth $\tau_{\rm photo}$ is unity for an observer at infinity. Its location $z_{\rm photo}$, measured from the equatorial plane, can be found by integrating over the wind density \citep{2009PASJ...61.1305F},
\begin{equation}\label{eq: wind photo}
    \tau_{\rm photo}=\int^\infty_{z_{\rm photo}} \gamma_{\rm w}(1-\beta_{\rm w} \cos{\Theta}) \, \kappa_{\rm co} \,\rho_{\rm co} {\rm d}z \approx \int^\infty_{z_{\rm photo}}  \kappa_{\rm co} \,\rho_{\rm co} {\rm d}z =1,
\end{equation}
where $\kappa_{\rm co}\approx \kappa$ is the opacity and $\rho_{\rm co}\approx \rho_{\rm w}$ (the subindex co refers to the comoving frame). Since we assumed a fully ionized wind, the opacity is dominated by free electron scattering ($\kappa=\sigma_{\rm T}/m_{\rm p}$, where $\sigma_{\rm T}$ is the Thomson scattering, and $m_{\rm p}$ is the proton mass). The location of the wind photosphere is completely determined by the mass-loss rate and thus by the accretion input of matter. The apparent photosphere of this wind is of interest for the calculation of the absorption of the radiation. 

Since the wind is optically thick, we assumed that it radiates as a blackbody. The temperature measured by an observer at infinity is given by \citep{2009PASJ...61.1305F}
\begin{equation}
     \sigma_{\rm T} T_{\rm w}^4\approx \dot{e} \, L_{\rm Edd}/4 \pi r^2,
\end{equation}
where $\dot{e}=0.1$ is the fraction of $L_{\rm Edd}$ that passes the comoving luminosity of the wind \citep{2022A&A...664A.178S}. The radiative power of the wind in supercritical sources is then limited to the Eddington luminosity \citep{2019ApJ...871..115Z}. The numerical results of the photospheric temperature and height for the three scenarios are detailed in Table \ref{tab: scenarios}. 

This wind ejected by the accretion disk interacts with the BHs orbiting the SMBH, but before studying this interaction, we present a description of the BH population in the next subsection.

\subsection{Cluster of BHs}\label{sec: cluster}

The SMBH dominates the orbital evolution of the objects within an influence radius given by \citep{2013ARA&A..51..511K}
\begin{equation}
    r_{\rm cluster} = G M_{\rm SMBH}/{\sigma_*^2},
\end{equation}
where $\sigma_*^2$ is the dispersion of the velocities of the orbital objects \citep{1972ApJ...178..371P}, and it is calculated from the relation \citep{2022MNRAS.515.3299G}
\begin{equation}
    \sigma_* = 2\times10^7 \, {\rm cm \, s^{-1}} ({M_{\rm SMBH}/3.097\times10^8 M_\odot)^{(1/4.384)}}.
\end{equation}
In our case, $\sigma_*=9.14\times10^6\,{\rm cm\,s^{-1}}$, so that $r_{\rm cluster}\approx1.65\times10^{19}\,{\rm cm}\approx 5.48\,{\rm pc}\approx 10^7\,r_{\rm g,SMBH}$. 

\begin{figure}
    \centering
    \includegraphics[width=9.5cm]{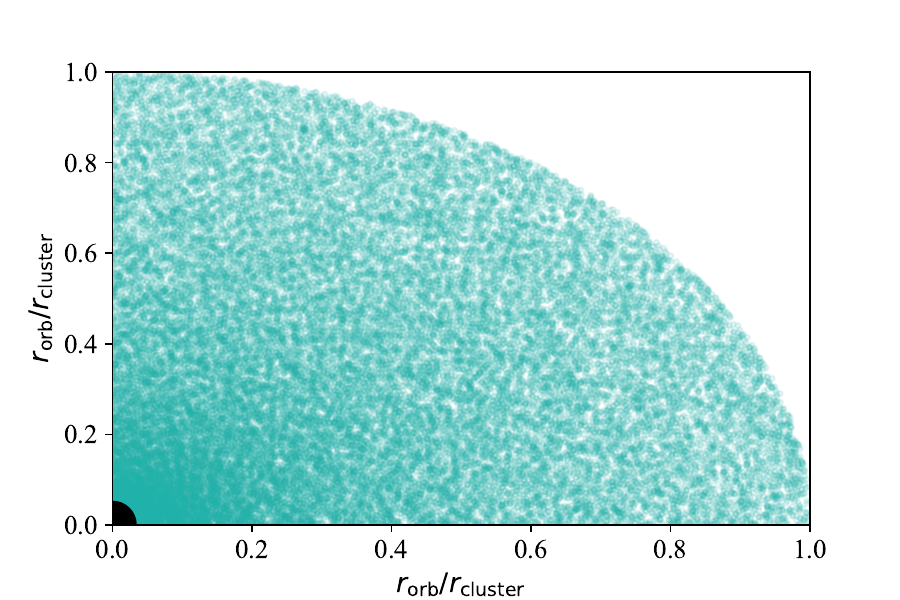}
    \caption{Distribution of BHs within the cluster projected on the $rz$-plane, where each dot represents a BH of $10M_{\odot}$. Both axes are on a linear scale and represent the distance $r_{\rm orb}$ to the SMBH in units of the cluster radius. The SMBH (large black dot, not to scale) yields $(0,0)$.}
    \label{fig: cluster}
\end{figure}

Most of the mass contained within this radius corresponds to stars and BHs. Each population is associated with a radial function of density, $\rho_{*}(r)$ and $\rho_{\rm BH}(r)$, where $r$ is the distance to the SMBH. The density profile of BHs is a fraction $C=(0.23M_{\odot}/ M_{\rm BH} )^{1/2}$ of the stellar profile \citep{2000ApJ...545..847M},\footnote{\cite{2019MNRAS.482.3669G} also give expressions for the density profiles of intermediate BHs with masses $10^2-10^3 M_{\odot}$ \citep{2009ApJ...697.1861A,2010ApJ...718..739M}  and $10^4 M_{\odot}$ \citep{2014ApJ...796...40M}. We will not consider BHs with such masses in this paper.}
\begin{equation}
    \rho_{\rm BH}(r)=C\rho_{\rm *}(r)=\left(\frac{0.23M_{\odot}}{M_{\rm BH}}\right)^{1/2} \frac{5}{16\pi}\frac{M_{\rm SMBH}}{r_{\rm cluster}^3}\left(\frac{r}{r_{\rm cluster}}\right)^{-7/4}.
\end{equation}
The cumulative mass distribution in a sphere with a radius $r$ is obtained by integrating $4\pi r^2\rho_{\rm BH}(r)$, which gives
\begin{equation}\label{eq: cumulative_mass}
    m_{\rm BH}(r)=\left(\frac{0.23M_{\odot}}{M_{\rm BH} }\right)^{1/2} M_{\rm SMBH}\left(\frac{r}{r_{\rm cluster}}\right)^{5/4},
\end{equation}
so the number of BHs inside the sphere is estimated to be $N_{\rm BH}(r) \sim m_{\rm BH}(r)/M_{\rm BH}$. 
We obtain a total number of $N_{\rm BH}\sim 1.5\times10^5$ isolated BHs that form the cluster for the assumed BH masses.\footnote{We note that \cite{2000ApJ...545..847M} calculated the number of BHs within a distance that is $\sim2.5$ times smaller than the influence radius of the SMBH they considered.} Figure \ref{fig: cluster} shows the distribution of BHs within the radius of the cluster projected on the $rz$-plane, where each dot represents an individual BH. 

\subsection{Accretion onto stellar black holes}
Stellar BHs orbiting the SMBH capture matter from the dense wind ejected by its accretion disk. Since the relation between the velocities of the wind and the BHs is not negligible, the accretion is not spherical, but cylindrical. That is, small accretion disks with radii $R_{\rm HL}=2GM_{\rm BH}/v_{\rm rel}^2$ are formed around the BHs via the BHL regime \citep[see section 4.3 of][for a full description of this regime]{Romero-Vila2014}. In the latter expression, $v^2_{\rm rel}= v^2_{\rm w} + v^2_{\rm BH}$ is the composition of the velocities of the spherical wind ejected from the SMBH disk and the BH in its orbital motion, $v^2_{\rm BH}=GM_{\rm SMBH}/r_{\rm orb}$. We assumed that the distance between the SMBH and the BH is constant, $r=r_{\rm orb}$ (circular Keplerian orbits). The outer radii of the accretion disks are in the range $R_{\rm HL}\sim 10^{4-5}\,r_{\rm g,BH}$ for BHs at different distances. 

The accretion disk is formed when the circularization radius ($R_{\rm circ}$) is larger than the radius of the innermost stable circular orbit of the BH ($R_{\rm ISCO}=6\,r_{\rm g,BH}$, for a nonrotating BH). The former is given by $R_{\rm circ}=J^2/GM_{\rm BH}$, where $J=R_{\rm HL}\,v_{\rm rel}$ is the angular momentum per unit mass of the wind when it is trapped in the gravitational field of the BH. We then have $R_{\rm circ} = R_{\rm HL}^2\,v_{\rm rel}^2/G\,M_{\rm BH} = 4G^2M_{\rm BH}^2\,v_{\rm rel}^2/G\,M_{\rm BH}\,v_{\rm rel}^4=4\,G\,M_{\rm BH}/v_{\rm rel}^2$. Hence, $R_{\rm circ}\approx 4\times 10^{27}\,{\rm cm^3\,s^{-2}} / v_{\rm rel}^2$ for the assumed value of the BH mass, which is greater than $R_{\rm ISCO}\sim9\times10^6\,{\rm cm}$ if $v_{\rm rel}<0.7\,c$ (this condition is always satisfied in this work). 

\begin{figure}
    \centering
    \includegraphics[width=9.5cm]{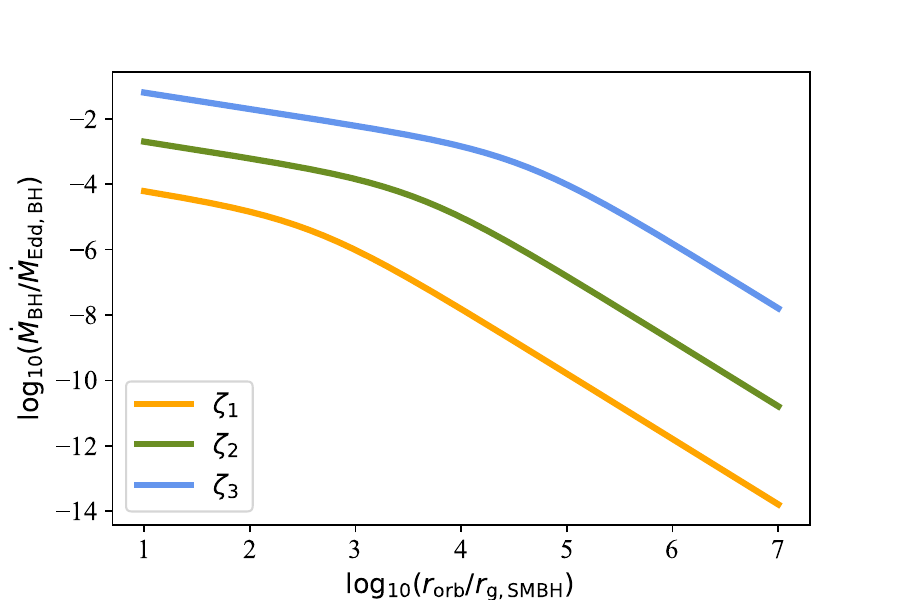}
    \caption{Accretion rates onto the isolated BHs orbiting the SMBH in units of their Eddington rate (see Eq. \ref{eq: acrecion BH}) as a function of their distance from the center of the galaxy. Both axes are on a logarithmic scale. We plot the three regimes of SMBH accretion ($\zeta_1 - \zeta_3$ from bottom to top).}
    \label{fig: accretion}
\end{figure}

The accretion rate of a BH by the BHL mechanism reads
\begin{equation} \label{eq: tasa_acc}
    \dot{M}_{\rm BH}=4\pi G^2 M_{\rm BH}^2 \rho_{\rm w} v_{\rm rel}^{-3}.
\end{equation}
Replacing the expressions for $\rho_{\rm w}$ (Eq. \ref{eq: densidad_viento}) and $v_{\rm rel}$ into the above equation, we obtain
\begin{equation}
    \dot{M}_{\rm BH}=G^2 M_{\rm BH}^2 \dot{M}_{\rm w}\,/\,r^2_{\rm orb}v_{\rm w}(v^2_{\rm BH}+v_{\rm w}^2)^{3/2}.
\end{equation}
The accretion disk of the SMBH is regulated at the Eddington rate, so that the mass-loss rate must be approximately equal to the accretion rate, $\dot{M}_{\rm w}\approx \dot{M}_{\rm input}$. Since the SMBH is supercritical, $\dot{M}_{\rm w}  = \zeta \, \dot{M}_{\rm Edd}$, with $\zeta\gg1$, where the Eddington rate is given by Eq. \ref{eq: tasa eddington}. Then, the accretion rate of the BH (with all quantities in cgs units) is
\begin{equation} \label{eq: acrecion BH}
    \dot{M}_{\rm BH}\approx 3.135\times10^{-30} \cdot \frac{\zeta \, M^2_{\rm BH} \, M_{\rm SMBH} }{r^2_{\rm orb} v_{\rm w}(v^2_{\rm w}+v^2_{\rm BH})^{3/2}} \ {\rm g\,s^{-1}}.
\end{equation}
The accretion rate onto the isolated BHs thus depends on four parameters: the superaccretion factor of the SMBH ($\zeta$), the BH masses ($M_{\rm BH}$ and $M_{\rm SMBH}$), and the location of the BHs ($r_{\rm orb}$). Therefore, when we fix the masses and the superaccretion factor, the accretion of matter onto the BHs only depends on the orbital radius. The accretion disk around the BH forms on the orbital plane, where the initial direction of the accreting material depends on the relation between the orbital velocity and the wind velocity. Figure \ref{fig: accretion} shows the accretion rate of stellar BHs on a logarithmic scale in units of the BH Eddington rate as a function of the orbital radius (in units of the gravitational radius of the SMBH) for the three different accretion rates onto the SMBH ($\zeta_1 - \zeta_3$: yellow, green, and light blue curves in the plot). 

We studied BHs inside ($r_{\rm orb}\le z_{\rm photo}$) and outside the photosphere ($r_{\rm orb} > z_{\rm photo}$). We note that in the case of scenario $\zeta_1$, almost the entire population of BHs is outside the photosphere. We computed the emission from the accretion disks only for the BHs outside the wind photosphere. In most cases, the accretion proceeds in the low-accretion ADAF (advection-dominated accretion flow) regime, with $\dot{M}_{\rm BH}< 10^{-3}\,\dot{M}_{\rm Edd}$. Since the disk contribution is small in any case, we assumed for simplicity that the disks radiate in the standard regime of \cite{1973A&A....24..337S}. This contribution should therefore be considered only as an upper limit.

\section{Jet} \label{sec: jet}

The BHs with their small accretion disks generate two jets that propagate in opposite directions through the wind of the supercritical SMBH \citep[see][]{2016A&A...590A.119B}. In this section, we first characterize the jet and its launch and then study the shocks at the base (recollimation) and the lobes (terminal region). We then discuss particle acceleration, cooling, and transport. Table \ref{tab: jet_particular} lists the values for the parameters of BHs near the wind photosphere for the three scenarios we considered.

\begin{table*}
\caption{Parameters of BHs and their jets.}
\centering
\begin{adjustbox}{width=18.4cm}
\begin{tabular}{cccccccccc}
\hline
\hline
\rule{0pt}{2.5ex} $\dot{M}_{\rm input}/\dot{M}_{\rm Edd,SMBH}$ & $r_{\rm orb}[{\rm cm}]$ & $v_{\rm orb}[{\rm cm\,s^{-1}}]$ & $\dot{M}_{\rm BH}[{\rm g\,s^{-1}}]$ & $L_{\rm j}[{\rm erg\,s^{-1}}]$ & $\rho_{\rm w}[{\rm g\,cm^{-3}}]$ & $z_{\rm recoll}[{\rm cm}]$ & $B_{0}[{\rm G}]$ & $B(z_{\rm recoll})[{\rm G}]$ & $l_{\rm j}[{\rm cm}]$ \\
\hline
\rule{0pt}{2.5ex} $\zeta_1$ & $3.75\times 10^{14}$ & $1.88\times 10^{9}$ & $8.4\times 10^{13}$ & $7.56\times 10^{33}$ & $5.28\times 10^{-14}$ & $1.9\times 10^{8}$ & $1.78\times 10^{5}$ & $1.37\times 10^{5}$ & $4.5\times 10^{14}$\\
$\zeta_2$ & $10^{16}$ & $3.64\times 10^{8}$ & $1.55\times 10^{14}$ & $1.39\times 10^{34}$ & $2.35\times 10^{-15}$ & $5.26\times 10^{9}$ & $2.56\times 10^{5}$ & $7203$ & $4.73\times 10^{14}$\\
$\zeta_3$ & $4.44\times 10^{17}$ & $5.49\times 10^{7}$ & $2.05\times 10^{14}$ & $1.84\times 10^{34}$ & $3.84\times 10^{-17}$ & $1.77\times 10^{11}$ & $2.95\times 10^{5}$ & $245$ &$4.34\times 10^{14}$ \\

\hline
\end{tabular}
\end{adjustbox}
\tablefoot{Parameters of BHs and their jets that contribute most to the emission (corresponding to those just outside the wind photosphere) for three accretion rates $\dot{M}_{\rm input}$ onto the SMBH ($\zeta_1\equiv 10\dot{M}_{\rm Edd}$, $\zeta_2\equiv 10^2\dot{M}_{\rm Edd}$, and $\zeta_3\equiv 10^3\dot{M}_{\rm Edd}$). All parameters are derived.} \label{tab: jet_particular}
\end{table*}

\subsection{Launch and general properties}
We assumed that the jet is perpendicular to the plane of the accretion disk and is fully formed at a distance of $z_0=100\,r_{\rm g}$ from the BH, with an initial radius of $r_0=0.1\,z_0=10\,r_{\rm g}$. Afterward, the outflow expands as a cone with radius $r_{\rm j}(z)=r_0(z/z_0)=0.05\,z$ \citep[i.e., with a semi-opening angle of $\theta_{\rm j}\approx 3^{\circ}$, see e.g.][]{2015A&A...584A..95P}. According to the disk-jet coupling hypothesis of  
\cite{1995A&A...293..665F}, the kinetic power of each jet is assumed to be proportional to the accretion power, $L_{\rm j}=q_{\rm j}L_{\rm accr}$, where $q_{\rm j}=0.1$ is a moderate efficiency in transferring power from the disk to the jet, and $L_{\rm accr}\approx \dot{M}_{\rm BH}\,c^2$. 
A fraction $q_{\rm rel}\approx0.1$ of the jet power is expected to be in the form of relativistic particles. Then we can write $L_{\rm rel}=q_{\rm rel}\,L_{\rm j}$. We included both the hadronic and leptonic fractions, $L_{\rm rel}=L_{\rm p}+L_{\rm e}$. 

We considered two acceleration regions mediated by shocks for relativistic particles in the jet: one near its base (recollimation), and the other at the terminal region (lobes).\footnote{We neglected other acceleration regions that may exist in the jet, such as internal shocks produced by the nonlaminar nature of the fluid in matter-dominated regions.} The energy distribution between hadrons and leptons is unknown; we assumed an equipartition near the base of the jet ($L_{\rm p}=\,L_{\rm e}$) and a hadronic-dominated scenario in the terminal region ($L_{\rm p}=100\,L_{\rm e}$) \citep[for a discussion about this assumption see e.g.,][]{2008A&A...485..623R, 2009A&A...497..325B}. We describe these regions in the following two subsections.




\subsection{Recollimation}

The recollimation of the jet produced by the wind of the SMBH determines the location of the acceleration region near the base of the jet. Given the orbital radii at which the BHs are located, we assumed that the fluid streamlines are parallel when the wind reaches the BHs.



The ram pressure of the wind at a height $z$ above  the BH is
\begin{equation}
    P_{\rm ram,w}=\rho_{\rm w}v_{\rm rel}^2=\frac{\dot{M}_{\rm w}v_{\rm rel}^2}{4\pi v_{\rm w} (r_{\rm orb}+z)^2}\approx \frac{\dot{M}_{\rm w}v_{\rm rel}^2}{4\pi v_{\rm w} r_{\rm orb}^2},
\end{equation}
where we considered the approximation $r_{\rm orb}\gg z$. On the other hand, the jet lateral pressure is \citep{2015ApJ...801...55Y,2016A&A...590A.119B}
\begin{equation}
    P_{\rm ram,j}=\rho_{\rm j}(v_{\rm j}\, \theta_{\rm j})^2 = \frac{L_{\rm j}(v_{\rm j}\, \theta_{\rm j})^2}{\pi (\theta_{\rm j}z)^2 \gamma_{\rm j}(\gamma_{\rm j}-1)c^2v_{\rm j}}=\frac{L_{\rm j}\beta_{\rm j}}{\pi z^2 \gamma_{\rm j}(\gamma_{\rm j}-1)c},
\end{equation}
where $v_{\rm j}$ is the bulk velocity of the flow. We assumed a moderate value for the Lorentz factor $\gamma_{\rm j}=(1-v_{\rm j}^2/c^2)^{-1/2}=3$, so $v_{\rm j}=0.943\,c$ \citep{2002A&A...390..751H}. We can find the height $z_{\rm recoll}$ of a recollimation shock in the jet by equaling these pressures, $P_{\rm ram,j}=P_{\rm ram,w}$,
\begin{equation}
    z_{\rm recoll}=[4L_{\rm j}\beta_{\rm j}v_{\rm w}r_{\rm orb}^2 / \gamma_{\rm j}(\gamma_{\rm j}-1)c \dot{M}_{\rm w}v_{\rm rel}^2]^{1/2}.
\end{equation}

The magnetic field at the base of the jet, $B_0=B(z_0)$, can be estimated from the equipartition between the magnetic and kinetic energy densities at $z_0$ ($\epsilon_{\rm B}=\epsilon_{\rm K}$), where the Alfv\'en radius is ($r_{\rm A}\equiv z_0$) and the geometry of $B$ starts to change from poloidal to toroidal,  $B_0^2/8\pi=L_{\rm j}/2\pi r_0^2 v_{\rm j}$. The magnetic field in the jet decreases with distance $z$ from the BH: when the flow expands adiabatically,
$B(z)=B_0(z_0/z)^m$, where $m$ is the magnetic exponent which depends on the geometry of $B$. In our case, $m=1$ because the magnetic field has a toroidal geometry for $z>z_0$. 

On the other hand, the number density of cold protons at a distance $z$ from the BH in the jet is $n_{\rm p}(z)\sim \dot{M}_{\rm j}/\pi\, r^2_{\rm j}\, m_{\rm p}\,v_{\rm j}$, where the mass-loss rate in the jet is given by $\dot{M}_{\rm j}=L_{\rm j}/\gamma_{\rm j}(\gamma_{\rm j}-1)c^2$.

\subsection{Terminal jet}

The head of the jet propagates through the wind. Its length $l_{\rm j}$ at a time $t$ is given by \citep{1997MNRAS.286..215K}:
\begin{equation}
    l_{\rm j}=(L_{\rm j}/\rho_{\rm w})^{1/5}\,t^{3/5}.
\end{equation}

We assumed that the lifetime of the single microquasar lasts for a fraction of the time that the SMBH spends in the super-Eddington regime (see Eq. \ref{eq: t edd}). 
We adopted conservative ages of $15$, $5$, and $1\,\rm yr$ for scenarios $\zeta_1$, $\zeta_2$, and $\zeta_3$, respectively. The jet lengths were about $l_{\rm j}\sim10^{14}\,{\rm cm}$ for all scenarios and orbital distances (the larger the orbital distance, the longer the jet). The jet length always exceeds the recollimation height, $l_{\rm j}\gg z_{\rm recoll}$ (see Table \ref{tab: jet_particular}). 

At the head of the jet, two shocks are generated: a forward radiative shock that propagates into the wind, and a reverse adiabatic shock that propagates into the jet. The radius of the bow shock is assumed to be $\sim l_{\rm j}/3$ \citep{2009A&A...497..325B}. Particle acceleration to relativistic energies only occurs in the reverse adiabatic shock \citep[see characterization of these shocks in][]{2023A&A...671A...9A}. 
Conversely, in radiative shocks, the gas emits thermal optically thin radiation, and no particle acceleration is produced \citep[see e.g.][]{2011hea..book.....L}. We assumed a magnetic field of $B\approx 10\,\mu{\rm G}$ for this region, which corresponds to that of the compressed medium \citep[the magnetic field can take maximum values of $\sim 100\,\mu{\rm G}$ in these scenarios, see][]{Araudo_etal_2015,Araudo_etal_2016}. 
The proton number density is $n_{\rm p}=\rho_{\rm w}/m_{\rm p}$.

\subsection{Relativistic particle acceleration, cooling, and transport}


The acceleration rate for a charged particle in a magnetic field $B$ through first-order Fermi mechanism is given by $t_{\rm acc}^{-1}=\eta_{\rm acc}\,e\,c\,B/E$, where $E$ is the energy of the particle, and $\eta_{\rm acc}$ is a parameter characterizing the efficiency of the acceleration mechanism (we assumed a mildly efficient acceleration in all regions, $\eta_{\rm acc}=0.1$). 

Particles accelerated at the shocks can cool by various processes and produce thermal and nonthermal radiation. The timescales associated with this cooling are related to the total energy loss of the particles: ${\rm d}E/{\rm d}t\approx -E/t_{\rm cool}$, where the total cooling rate is $t_{\rm{cool}}^{-1} = \sum_i t_{i}^{-1}$ ($t_i$ corresponds to each timescale of the cooling processes involved).

We assumed advective escape at the base of the jet and diffusive escape at the lobes. When the cooling timescales are shorter than the escape timescales, particles radiate before escaping from the acceleration regions. The maximum energy for each type of particle can be derived by equating the acceleration rate to the total cooling or escape rate. This energy cannot exceed the maximum energy imposed by the Hillas criterion. 

Radiative cooling is caused by nonthermal processes as a consequence of the interaction of particles with ambient fields and matter.  
Our model is lepto-hadronic, and we therefore numerically calculate the following cooling processes:

--Synchrotron: Interaction of protons and electrons with the ambient magnetic field.

--Inverse Compton: Collision of relativistic electrons with photons of the ambient radiation field.

--Self-synchrotron Compton (SSC): Collision of relativistic electrons with photons of the synchrotron radiation field.

--Relativistic Bremsstrahlung: Coulombian interactions between relativistic electrons and cold matter.

--Thermal Bremsstrahlung: Free-free radiation by Coulombian interactions. 

--Photo-hadronic interactions: Interaction of highly relativistic protons with photons of the surrounding radiation field.  This interaction produces pions that decay into gamma rays, $e^+e^-$ pairs, and neutrinos.

--Proton-proton: Collision of relativistic protons with cold matter. This process is also a source of gamma rays, pairs, and neutrinos.


The relativistic particles have a distribution given by ${\rm d}N= n(\vec{r},E,t){\rm d}E{\rm d}V$, where $n$ is the particle number density, $t$ is the time, $\vec{r}$ is the position, $V$ is the volume, and $E$ is the energy. The transport equation governs the evolution of this distribution \citep{1964ocr..book.....G}. We solved this equation numerically in steady state and in the one-zone approximation and assumed an injection spectral index for the particles of $p\sim 2$ in all cases \citep[e.g.,][]{1983RPPh...46..973D,2008A&A...485..623R,2015A&A...584A..95P}.

After we determined the relevant processes involved in the cooling and calculated the particle distributions, we calculated the SED of the radiation produced by the particles as they interact with the magnetic, matter, and radiation fields (we refer to \cite{2008A&A...485..623R,Romero-Vila2014} and references therein for additional details on nonthermal processes and to \cite{2011hea..book.....L} for details on thermal Bremsstrahlung (see Chapter 6.5)).

\section{Results} \label{sec: results}

\begin{figure}
  \centering
  \begin{minipage}{0.5\textwidth}
    \centering
    \includegraphics[width=9.3cm]{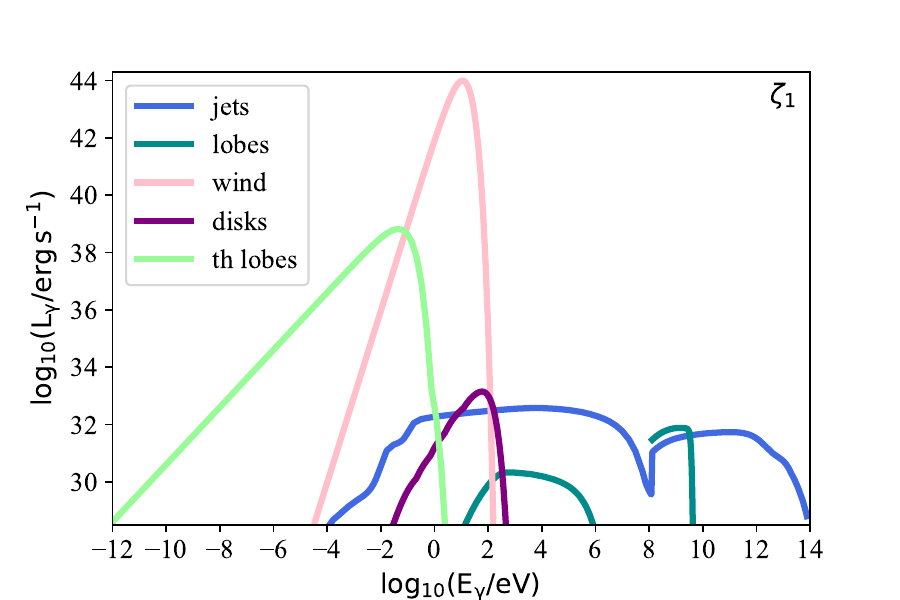}
  \end{minipage}
  \hfill
  \begin{minipage}{0.5\textwidth}
    \centering
    \includegraphics[width=9.3cm]{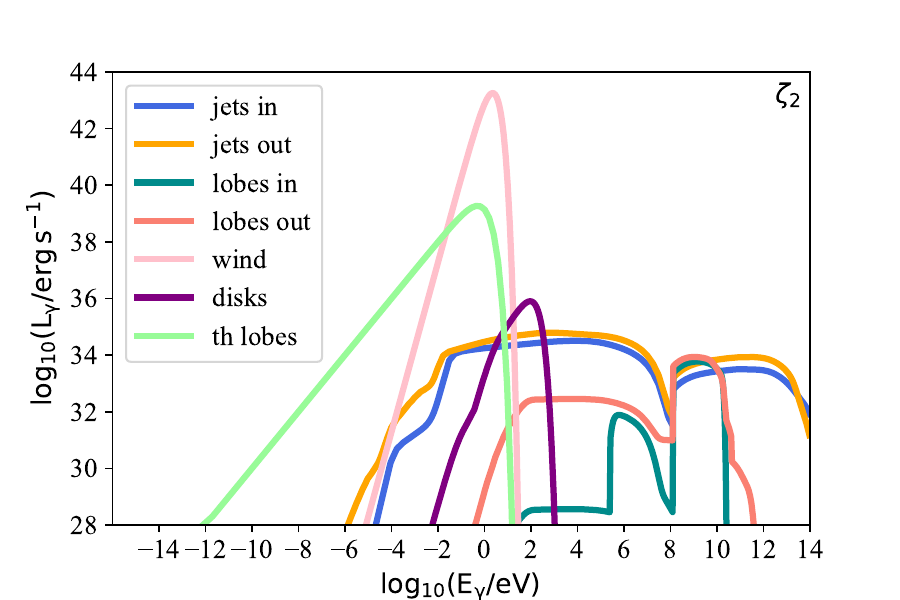}
  \end{minipage}
  \hfill
  \begin{minipage}{0.5\textwidth}
    \centering
    \includegraphics[width=9.3cm]{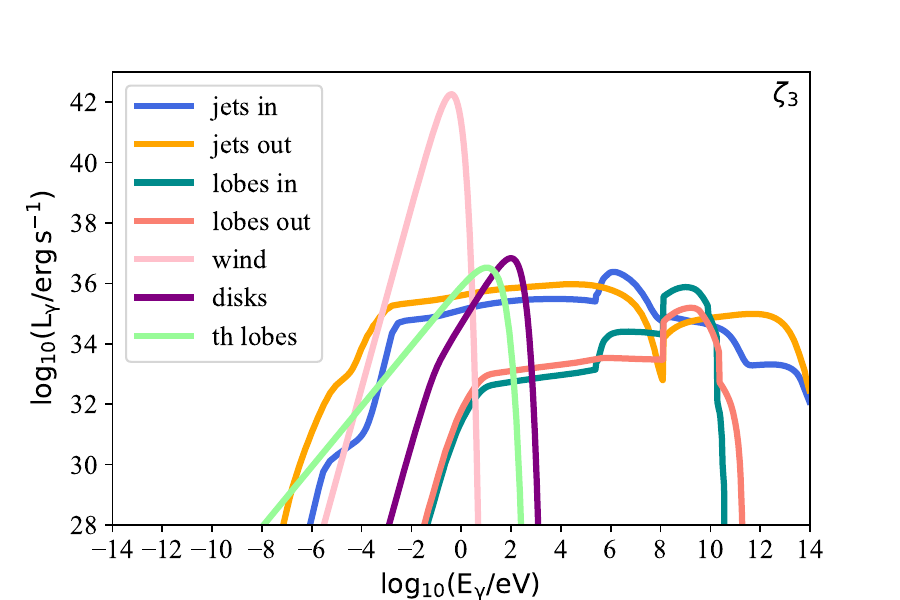}
  \end{minipage}
  \caption{Emission of the entire system for scenarios $\zeta_{1}$, $\zeta_{2}$, and $\zeta_{3}$, from top to bottom. We plot the nonthermal SEDs of the base (jets) and terminal jets (lobes) for two groups of BHs, those inside (in) and outside (out) the photosphere (except for the case $\zeta_1$, where we present only the results for BHs outside the photosphere). We also plot the  SEDs of the SMBH wind photosphere, the accretion disks of the BHs, and the thermal lobes (th lobes) outside the photosphere.}
  \label{fig: total sed results}
\end{figure}

\begin{figure}
    \centering
    \includegraphics[width=9.3cm]{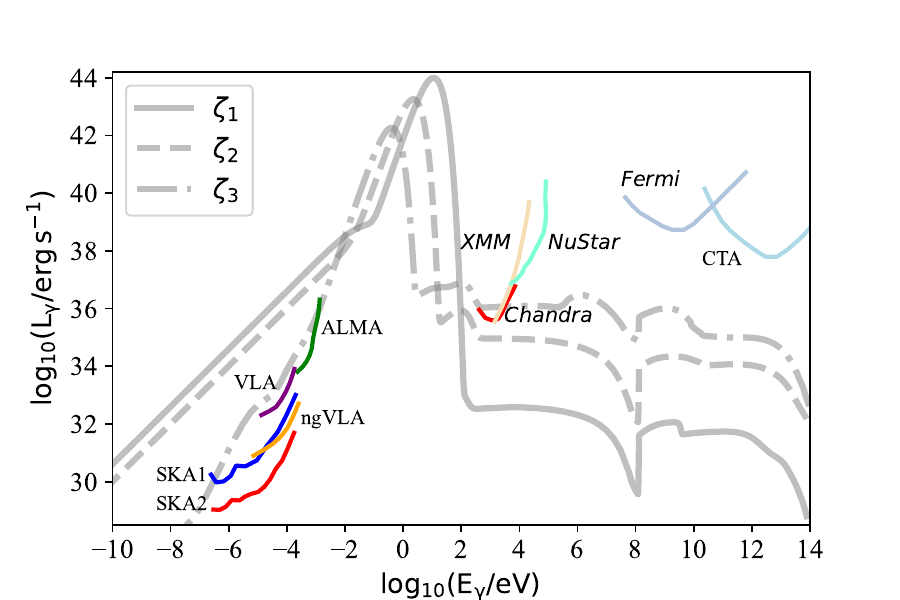}
    \caption{Total SEDs of the three scenarios and sensitivity curves of several ground- and space-based telescopes, assuming a distance to the source of $\sim3\,{\rm Mpc}$. We plot with solid, dashed, and dot-dashed lines the total SEDs of scenarios $\zeta_1$, $\zeta_2$, and $\zeta_3$. The sensitivities of ALMA, VLA, and ngVLA are adapted from \cite{2020ApJ...901...39G} (1 h on-source integration); the SKA1 and SKA2 sensitivities are taken from \cite{SKA_sensitivity} (1 h); the sensitivities of \textit{XMM-Newton} ($10^5\,$s), \textit{Chandra} ($10^5\,$s), \textit{NuStar} ($10^6\,$s), \textit{Fermi} ($10\,$yr), and CTA ($50\,$h) are taken from \cite{2022JCAP...08..013L} and references therein (the integration time for observations is given in brackets).}
    \label{fig: seds&sensitivities}
\end{figure}

The results for the accretion onto the BHs and their location in the cluster (along with the properties of the supercritical SMBH wind) were presented in Sect. \ref{sec: model}. In this section, we focus on the results for the radiation produced by the system.

As mentioned in Sect. \ref{sec: cluster}, we applied our model to a population of $\sim1.5\times 10^5$ BHs with typical masses of $10\,M_{\odot}$ orbiting a $10^7\,M_{\odot}$ SMBH. The SED of the entire population of BHs was estimated as follows. First, we calculated the luminosity of individual BHs at different orbital radii BH$(i)$, with accretion rates according to the results shown in Fig. \ref{fig: accretion}. The total radiative contribution $L_{{\rm BH}(i)}$ of a single microquasar corresponds to the thermal emission from its mini-disk $(L_{\rm disk})$, the nonthermal emission from the recollimation shock $(L_{\rm recoll})$, the nonthermal emission from the reverse shock of its jet $(L_{\rm lobe,NT})$, and the thermal emission from the forward shock in the lobes $(L_{\rm lobe,thermal})$: $L_{{\rm BH}(i)}(E_\gamma)=L_{{\rm disk}(i)}(E_\gamma)+L_{{\rm recoll}(i)}(E_\gamma)+L_{{\rm lobe,NT}(i)}(E_\gamma)+L_{{\rm lobe,thermal}(i)}(E_\gamma)$. The emission produced at the recollimation shock was calculated taking relativistic effects into account: We calculated the Doppler boost factor $D^2=(\gamma_{\rm j}-\beta_{\rm j}\cos{\alpha_{\rm j}})^{-1}$, and we then average the emission over $\alpha_{\rm j}$ from $0^{\circ}$ to $90^{\circ}$. When $L'_{\rm recoll}(E_\gamma)$ is the comoving emission, then the boost luminosity in the observer's frame is $L_{\rm recoll}(E_\gamma) = \overline{D}^2\,L'_{\rm recoll}(E_\gamma)$ \citep{1985ApJ...295..358L}.

The second step was to calculate the gamma absorption by pair creation from photon-photon annihilation. 
We considered two radiation fields: the synchrotron emission produced in the jet $(a)$, and photons of the ambient radiation field produced by the wind photosphere $(b)$. Since the wind is dense, we also calculated the absorption by Thomson scattering $(c)$ for the BHs inside the photosphere. The attenuation factor is given by ${\rm{e}}^{-\tau(E_{\gamma})}$, where $\tau=\tau_a + \tau_{b} + \tau_{c}$ is the optical depth of the medium, which quantifies the absorption of radiation by all the processes involved. The attenuated luminosity is then $L_{{\rm BH}(i)}'(E_\gamma)=L_{{\rm BH}(i)}(E_\gamma)\cdot {\rm e}^{- \tau(E_{\gamma})}$.

Finally, we set up spherical shells centered on the SMBH for the different orbital radii considered so far, calculated the number $N_{{\rm BH}(i)}$ of BHs contained therein, and estimated the luminosity of the shell, assuming that all BHs produce approximately the same radiation (the shell is thin enough to guarantee homogeneity conditions): $L_{{\rm shell}(i)}(E_\gamma)=N_{{\rm BH}(i)}\, L_{{\rm BH}(i)}(E_\gamma)$. The total SED of the cluster is the sum of all contributions, $L_{\rm total}(E_\gamma)=\sum_{i} L_{{\rm shell}(i)}(E_\gamma)$. 

In what follows, we first detail the results for the emission from one single microquasar, and we then present the results for the emission from the entire population of BHs, plus the thermal contribution from the wind photosphere.



\subsection{Emission from individual sources}

All the thermal and nonthermal processes ultimately depend on the accretion rate onto the BHs and thus on the distance to the SMBH. A higher accretion rate provides more energy to transfer power to the jet, so that the major radiative contributions are produced by the BHs that are located just above the photosphere of the SMBH wind, $r_{\rm orb}\gtrsim z_{\rm photo}$. The length of the jet is also a crucial parameter for the thermal emission of the lobes. 

The emission of BHs inside the photosphere is entirely suppressed for photon energies $\lesssim 1\,{\rm MeV}$, except for those close to the photosphere border, whose emission is partially attenuated and provides the radiation that can escape from inside the photosphere and dominates the emission for that group of BHs. Below, we describe the results we obtained for the different parts of the single microquasar for the three scenarios we considered. In scenario $\zeta_1$ almost the whole population of BHs is outside the photosphere and we therefore did not calculate the contribution from the inner BHs, while in scenarios $\zeta_2$ and $\zeta_3$, we ruled out the thermal emission produced inside the photosphere because it is totally suppressed. Since the outer radii of the accretion disks are limited by the BHL radius, their spectra resemble a blackbody and not a characteristic multicolor disk. 

In what follows, we describe the radiative output for the different emission regions of the individual sources in the three scenarios (Fig. \ref{fig: total sed results} shows the results for the entire population described in the next subsection, not for the individual sources).

\noindent Scenario $\zeta_1$:

-- Accretion disks: The emission peaks at ultraviolet (UV) energies for the BHs near the photosphere and at optical energies for those in the outer cluster. The luminosity range is $\sim10^{25-33}\,{\rm erg\,s^{-1}}$, where the disks with a maximum in the UV are more powerful. 
    
-- Recollimation shock: The recollimation height is $\sim10^{8}\,{\rm cm}$ for the BHs that lie close to the photosphere. The magnetic field there is $\sim 10^5$G, and the dominant cooling process for accelerated electrons is synchrotron in all cases, reaching maximum energies of $10^{11}\,{\rm eV}$. The advective escape dominates in all cases for protons, reaching energies in the range $10^{14}-10^{15}\,{\rm eV}$. The emission is completely dominated by the BHs that lie close to the photosphere, producing synchrotron radiation with a luminosity  $\sim10^{31}\,{\rm erg\,s^{-1}}$ in the range $10^{-1}-10^{7}\,{\rm eV}$, and $\pi^0$-decay gamma emission due to $pp$ interaction with a luminosity above $\sim10^{30}\,{\rm erg\,s^{-1}}$ in the range $10^{8}-10^{12}\,{\rm eV}$. 

-- Nonthermal lobe: Electrons mainly cool by IC and Bremsstrahlung, reaching energies in the range $10^{8}-10^{11}\,{\rm eV}$. Protons reach their maximum energy of $10^{12}\,{\rm eV}$ in the lobes of BHs just above the photosphere, with $pp$ being the dominant cooling mechanism. The SEDs of the BHs are dominated by the IC and $pp$ emission, with luminosities for the BHs near the photosphere of $\sim10^{30}$ and $\sim10^{32}\,{\rm erg\,s^{-1}}$, respectively. Photon-photon annihilation is irrelevant.

-- Thermal lobe: The emission is completely dominated by the BHs near the photosphere, with a maximum luminosity of $\sim10^{39}\,{\rm erg\,s^{-1}}$ at $10^{-2}\,{\rm eV}$. This is the dominant emission mechanism for scenario $\zeta_1$.

\noindent Scenario $\zeta_2$:

-- Accretion disks: The emission peaks at soft X-ray energies for the BHs near the photosphere and at UV energies for those in the outer cluster. The luminosity range is $\sim10^{28-34}\,{\rm erg\,s^{-1}}$, where the disks with a maximum at X-rays are more powerful. 

-- Recollimation shock: The recollimation height is in the range $\sim10^{9}-10^{11}\,{\rm cm}$ above the BHs. The magnetic field there for the BHs outside the photosphere is $\sim 10-10^4$G, and the dominant cooling process for the accelerated electrons is synchrotron in all cases, reaching energies in the range $10^{11}\,{\rm eV}$ (near the photosphere) and $10^{13}\,{\rm eV}$ (near the cluster radius). The advective escape dominates in all cases for protons reaching energies in the range $10^{12}-10^{15}\,{\rm eV}$. The emission is completely dominated by the BHs that lie close to the photosphere, producing synchrotron radiation with a luminosity  $\sim10^{31}\,{\rm erg\,s^{-1}}$ in the range $\sim 10^{-2}-10^{7}\,{\rm eV}$, and $\pi^0$-decay gamma emission due to $pp$ interaction with a luminosity $\sim10^{30}\,{\rm erg\,s^{-1}}$ in the range $10^{8}-10^{12}\,{\rm eV}$. Internal and Thomson absorption is almost total for the BHs inside the photosphere and is not relevant for those outside it.

-- Nonthermal lobe: Electrons mainly cool by IC and Bremsstrahlung, reaching energies in the range $10^{9}-10^{12}\,{\rm eV}$. The maximum energy of the protons is $10^{12}\,{\rm eV}$; diffusive escape dominates in all cases, with $pp$ being the dominant cooling mechanism. The SEDs of the BHs outside the photosphere are dominated by IC and $pp$ emission, with luminosities of $\sim10^{30}$ and $\sim10^{32}\,{\rm erg\,s^{-1}}$ for the BHs outside but close to the photosphere. Photon-photon annihilation is only relevant for BHs inside the photosphere.

-- Thermal lobe: The luminosity range is $\sim10^{26}-10^{37}\,{\rm erg\,s^{-1}}$, and in all cases, the emission peaks at $1\,{\rm eV}$. This is the dominant emission mechanism for scenario $\zeta_2$.

\noindent Scenario $\zeta_3$:

-- Accretion disks: The emission peaks at soft X-ray energies for the BHs near the photosphere and at UV energies for those in the outer cluster. The luminosity range is $\sim10^{31-34}\,{\rm erg\,s^{-1}}$, where the disks with a maximum in X-rays are more powerful. This is the dominant emission mechanism for scenario $\zeta_3$.

-- Recollimation shock: The recollimation height is in most cases $\sim10^{11}\,{\rm cm}$ above the BHs. The magnetic field there for the BHs outside the photosphere ranges from $\sim 10-100\, {\rm G}$ and the dominant cooling process for the accelerated electrons is synchrotron in all cases, reaching energies in the range of $10^{10}\,{\rm eV}$ (near the photosphere) and $10^{13}\,{\rm eV}$ (near the cluster radius). The advective escape dominates in all cases for protons, reaching energies in the range $10^{14}-10^{15}\,{\rm eV}$. Again, the emission is completely dominated by the BHs that lie close to the photosphere, producing synchrotron radiation with a luminosity above $\sim10^{31}\,{\rm erg\,s^{-1}}$ in the range $10^{-3}-10^{5}\,{\rm eV}$, and $\pi^0$-decay gamma emission due to $pp$ interaction with a luminosity above $\sim10^{30}\,{\rm erg\,s^{-1}}$ in the range $10^{8}-10^{12}\,{\rm eV}$. As in the case of $\zeta_2$, internal and Thomson absorption is almost total for the BHs inside the photosphere and negligible for those outside.

-- Nonthermal lobe: Electrons mainly cool by IC and Bremsstrahlung, reaching energies in the range $10^{9}-10^{11}\,{\rm eV}$. Protons reach their maximum energy of $10^{11}\,{\rm eV}$ in the lobes of the BHs just above the photosphere, while the $pp$ interaction suppresses the acceleration for those close to the SMBH. Diffusive escape dominates in all cases, with $pp$ being the dominant cooling mechanism. The SEDs of the BHs outside the photosphere are dominated by the IC and $pp$ emission, with luminosities of $\sim10^{30}$ and $\sim10^{32}\,{\rm erg\,s^{-1}}$ for the BHs near the photosphere. Photon-photon annihilation is only relevant for the BHs inside the photosphere.

-- Thermal lobe: Although all thermal lobes are about the same size, the spectrum is more luminous for the BHs that lie close to the photosphere because the density of the medium is higher. The luminosity ranges from $\sim10^{28}-10^{33}\,{\rm erg\,s^{-1}}$, and in all cases, the emission peaks at $10\,{\rm eV}$.

\subsection{Emission from the entire system}

Figure \ref{fig: total sed results} shows the thermal and nonthermal radiative contributions of the whole system for scenarios $\zeta_1$, $\zeta_2$, and $\zeta_3$ (from top to bottom). We separated the nonthermal emission into two groups: the BHs inside and outside the photosphere of the SMBH wind (except for case $\zeta_1$, where we present only the results for the BHs outside the photosphere). The wind produces the peak contribution in all scenarios, with luminosities $\sim10^{42}-10^{44}\,{\rm erg\,s^{-1}}$ in the infrared. The thermal emission from the lobes of the jets is the dominant radiation produced by the clusters of scenarios $\zeta_1$ and $\zeta_2$. Conversely, in the $\zeta_3$ scenario, the dominant emission of the cluster is in the X-ray band and is mainly produced by the electron synchrotron mechanism at the base of the jets outside the photosphere, reaching a luminosity of $\sim10^{36}\,{\rm erg\,s^{-1}}$ for the entire population of BHs. The accretion disks also contribute at very soft X-rays with $\sim10^{37}\,{\rm erg\,s^{-1}}$. 

The very high energy emission is dominated by the $\pi^0$-decay due to the $pp$ interaction at the base and the lobes.  The emission from the base of the jet dominates that from its terminal region in all cases and energies, except for the $10^8-10^{10}\,{\rm eV}$ energy range, where the luminosity caused by $pp$ interactions from the lobes is slightly dominant.

Figure \ref{fig: seds&sensitivities} shows the total SEDs of the three scenarios and the sensitivities of different space- and ground-based telescopes for a distance of $3\,{\rm Mpc}$ corresponding to a nearby galaxy. In scenarios $\zeta_1$ and $\zeta_2$, the emission from a BHs cluster in a nearby galaxy is potentially detectable only at energies in the range $10^{-7}-10^{-4}\,{\rm eV}$ by the ground-based instruments Atacama Large Millimeter Array (ALMA), Very Large Array (VLA), next-generation VLA (ngVLA), and Square Kilometer Array (SKA1 and SKA2), although in the case of ALMA, the emission would probably be difficult to separate from that coming from the SMBH wind. In scenario $\zeta_3$, the electron synchrotron emission at such low energies could only be detected by ngVLA or SKA2. In the X-ray band, the \textit{XMM-Newton} and \textit{Chandra} telescopes might both also detect the synchrotron emission of the recollimation shock (although the detection margin is narrow with the parameters assumed in our model), while the \textit{NuStar} telescope is not able to detect the emission in any case.

The gamma-ray emission predicted by our model is not detectable by any of the current instruments, whether ground-based like Cherenkov Telescope Array (CTA) or space-based like \textit{Fermi}.

\section{Discussion} \label{sec: discussion}


\subsection{Dependence of the emission on the parameters}

In all scenarios and for all emission regions in the single microquasars, the radiative output is strongly dominated by the BHs near the photosphere. The size of the latter depends on the accretion rate onto the SMBH (the higher the rate, the larger the photosphere). The emission from a small number of BHs in the central parsec then dominates the remaining population within the gravitational influence radius $r_{\rm cluster}$. We note that the dense medium in which the BHs orbit can induce a migration to closer orbits over long periods of time, so that perhaps the population of BHs near the photosphere is underestimated in our model, and therefore, the emission is higher than predicted.

We have applied our model to conservative values for the BH masses: $M_{\rm SMBH}=10^7\,M_{\odot}$ and $M_{\rm BH}=10\,M_{\odot}$. Since $v^2_{\rm BH}\propto M_{\rm SMBH}^{1/2}$, then $\dot{M}_{\rm BH}\propto M_{\rm SMBH}^{-1/2}$ (see Eq. \ref{eq: acrecion BH}), and thus, a lower SMBH mass leads to a higher accretion rate onto the BHs. In addition, for the BHs we have $\dot{M}_{\rm BH}\propto M_{\rm BH}^{2}$, so that when we consider $M_{\rm BH}=30\,M_{\odot}$ and $M_{\rm SMBH}=10^6\,M_{\odot}$, then $\dot{M}_{\rm BH}$ increases by an order of magnitude. The accretion onto BHs is even more accentuated in the case of dwarf galaxies with $M_{\rm SMBH}\sim 10^5\,M_{\rm \odot}$ \citep[see e.g.][]{2015ApJ...809L..14B}. Since $L_{\rm j}\propto \dot{M}_{\rm BH}$ and the radiative output of the system is $\propto L_{\rm j}$, the luminosity of the cluster can increase up to two orders of magnitude in a galactic center with these masses. The accretion rate onto the stellar BHs is also strongly sensitive to the velocity of the SMBH wind, since $\dot{M}_{\rm BH}\propto v_{\rm w}^{-3}$. When instead of a smoothed outflowing wind, we consider inhomogeneities or retarding effects due to internal interactions, then its velocity may decrease \citep{ShoreSteveBook1992}. For example, reducing the wind speed by half would increase the accretion $\dot{M}_{\rm BH}$ by a factor of 10.

We did not consider the possible rotation of the BHs. For spinning BHs, the disk-jet coupling factor $q_{\rm j}$ can be larger than 0.1, leading to more powerful jets \citep{2011MNRAS.418L..79T,2012MNRAS.423.3083M}. Furthermore, the innermost stable circular orbit may be smaller than $6\,r_{\rm g,BH}$ (reaching $1\,r_{\rm g,BH}$ for maximally rotating BHs), so that the thermal emission from the disks will have its maximum at higher energies, favoring detection by X-ray telescopes in scenario $\zeta_3$. We note, however, that the Blandford–Znajek mechanism can be activated in spinning BHs under certain circumstances, producing jets without the formation of an accretion disk, as shown by \cite{2012MNRAS.421.1351B}. A harder spectral index for the injection of particles at the base of the jet would also increase the X-ray luminosity. On the other hand, a lower Lorentz factor of the jet would lead to stronger emission from the recollimation shock in the observer's frame, which could be higher by up to an order of magnitude. 



\subsection{Detectability}


The rate of TDEs is theoretically estimated to be $10^{-4}\,{\rm yr^{-1} gx^{-1}}$  \citep{2016MNRAS.455..859S}, although other authors suggested higher rates, from $10^{-3}{\rm yr^{-1} gx^{-1}}$ \citep{kaur2024elevated} to $10^{-2}{\rm yr^{-1} gx^{-1}}$ \citep{Tadhunter_etal_2017_nature}. These theoretical predictions differ from observations, the rate of which is $\sim10^{-5}\,{\rm yr^{-1} gx^{-1}}$. \cite{2016MNRAS.455..859S} stated that selection effects, small number statistics, and the possible influence of dust or photoelectric extinction could all contribute to this discrepancy between theory and observations (TDEs are difficult to observe because they have their characteristic emission at optical frequencies, and the galactic centers are heavily obscured by dust).


In transient events, jet lengths are strongly constrained. Instead, a persistent super-Eddington scenario with typical microquasar ages of $10^5$yr leads to jet lengths on the order of $0.1-1$pc. In this case, the terminal lobes of the jet (both the reverse- and forward-shock regions) will be much larger, since the size of the lobes is proportional to the length of the jet, so that the emission regions and luminosity will be larger as well. The thermal emission in cases $\zeta_1$ and $\zeta_2$ would be higher, and a larger acceleration region of the reverse shock in scenario $\zeta_3$ will also allow the particles to cool by the synchrotron mechanism, favoring radio emission in the SKA range.

The least likely scenario to be detected is $\zeta_3$ because of the very high accretion rate and the duration of the TDE in this case. Its detection should occur in the X-ray band and could only happen for nearby galaxies. Conversely, BHs of scenarios $\zeta_1$ and $\zeta_2$ could be detected at very low energies with the ngVLA and SKA telescopes from a distance up to 100 Mpc in the case of SKA1 and SKA2. 

The emission from X-ray binaries orbiting the SMBH would probably exceed the cluster emission in the X-ray band. When we compare our results with Cyg X-1 \citep[see][]{2015A&A...584A..95P}, for example, the emission from isolated BHs dominates in the radio bands by many orders of magnitude.

\subsection{Further considerations}


The effect of the strong wind on the jet can lead to jet-bending effects, as detailed, for example, by \cite{2015ApJ...801...55Y} and \cite{2016MNRAS.456.3638Y} for the case of X-ray binaries.  \cite{2016A&A...590A.119B} also investigated orbital effects on the jet, such as the Coriolis effect, which in certain cases can lead to a helical structure of the jet trajectory \citep[see also][]{2020A&A...638A.132B}. We neglected these effects in this work, although they may be relevant for some of the BHs in the cluster under certain conditions, especially the bending (the Coriolis force affects the system on scales much larger than the jet lengths discussed in this paper).  

In a TDE, the wind is produced during the super-Eddington phase. The duration of this stage in the scenario $\zeta_3$ is estimated to be $\sim 1\,$yr, and the wind has a velocity of $\sim 10^8 \,{\rm cm\, s^{-1}}$. Since the BHs that contribute most to the radiation are located at $\sim 4\times 10^{17}$cm, just above the wind photosphere, it takes $\sim 30\,$yr for the wind to reach these BHs, which is much longer than the super-Eddington phase. In these cases, the outflow from the SMBH resembles a dense shell illuminating the single BHs from the cluster at different times (first those closer to the SMBH, then those farther away). Therefore, there could be a delay of years between the emission of the TDE from the core of the system and the radiative activation of the BHs. 

After the tidal disruption of the star, the wind launched by the transient accretion disk formed around the SMBH propagates as an expanding shell with a constant thickness $\delta x$, so that at some point it becomes transparent to the inner radiation. This can be roughly estimated by imposing $\tau = \sigma_{\rm T}\, \delta x\, n(d) \lesssim 1$, where $\tau$ is the optical depth, $\sigma_{\rm T}$ is the Thomson cross-section, and $n$ is the number density of the wind at a certain distance $d$ from the SMBH.  For example, for the parameter values in the scenario $\zeta_3$, the super-Eddington phase lasts $t\sim1\,$yr, and the thickness of the wind shell therefore is $v_{\rm w}\,t\sim 10^{15}\,{\rm cm}$. At an orbital radius $z_{{\rm photo}, \zeta_3}/2\sim 10^{17}\,$cm (see Table \ref{tab: scenarios}), the ambient number density is $\sim3\times 10^8\,{\rm cm^{-3}}$, so that $\tau<1$ and the inner radiation passes through the wind. The implications of this situation will be addressed in future work.

In TDEs, the emission from the AGN is produced during the lifetime of the disk. First, the accretion disk forms, producing a multiband spectrum. Some time later, the winds are launched and screen the emission from the disk. Radio emission could be present when the system has a jet, but we assumed a nonjetted SMBH \citep[][suggesting that only a fraction of 0.3 percent of TDEs produce jets]{2016MNRAS.455..859S}. In addition, jet activity is significantly shorter than the lifetime of the accretion disk formed during a TDE \citep[e.g.,][]{2015MNRAS.450.2824M,Zubovas_2019}. Furthermore, if there is a delay between the TDE emission and the illumination of the individual BHs in the cluster, as we suggested above, the transient active galaxy turns off as the BHs turn on. This could also aid in the detection of the BHs because the primary event will alert us to the possibility that the mechanisms discussed in this paper are activated.

The terminal region of the jets injects cosmic rays with energies up to $\sim 1$ TeV. The relativistic protons and electrons escaping from the lobes interact with the surrounding medium, producing diffuse emission in the surroundings through $pp$ and Bremsstrahlung interactions. In addition, when protons escape from the central region, they may interact with cold matter at larger distances, producing gamma rays. Neutrons created at the recollimation shock via $p+\gamma \rightarrow n + \pi$ can escape from the jet and interact with the dense surrounding gas, also leading to the production of gamma rays \citep{2021A&A...650A.136E}.

The jets of the stellar BHs orbiting close to the SMBH can also interact with the super-Eddington disk. This interaction can lead to particle acceleration, nonthermal emission, and neutrinos under certain conditions. Neutrino emission can play some role in the detection of these systems, since the wind photosphere is transparent to them. The interaction of jets with dense targets of matter was addressed, for example, by \cite{2005A&A...439..237R}, who studied misaligned jets interacting with the atmospheres of stellar companions in X-ray binary systems. Neutrino production from cosmic ray-disk interactions was addressed by \cite{2003ApJ...589..481A}. We plan to investigate the BH jet-SMBH disk interaction in detail in future work (see \citealt{Tagawa_etal2023b,Tagawa_etal2023a} for preliminary studies of the high-energy electromagnetic, neutrino, and cosmic-ray emission by stellar mass BHs inside disks of AGNs).



\section{Conclusions}\label{sec: conclusions}

We have developed a model for exploring the interaction between the wind of super-Eddington SMBHs and clusters of stellar-mass BHs located at galactic centers. The BHs accrete matter from the dense wind and produce jets. We applied our model to an SMBH of $10^7M_{\odot}$ and a cluster of $\sim 10^5$ BHs of $\sim 10M_{\odot}$, and we studied three accretion regimes of 10, $10^2$, and $10^3$ Eddington rates. We found that both thermal and nonthermal radiation are produced by the BHs of the cluster, which become "single"  microquasars that formed as a consequence of the accretion of matter from the SMBH wind. The total emission produced by the cluster would last as long as the system is in the super-Eddington regime and may be observable at X-ray energies (with \textit{Chandra} and \textit{XMM-Newton}) and radio wavelengths (e.g., with VLA and SKA) in nearby galaxies, providing a tool for probing the central population of stellar black holes.    


\begin{acknowledgements}
This work was supported by grant PIP 0554 (CONICET). The authors thank the referee for his/her useful comments that improved this work. LA thanks Florencia Rizzo, Jiri Hor\'ak, Vladimir Karas, and Pablo Sotomayor for useful discussions, and the UNLP for the education received. GER acknowledges financial
support from the State Agency for Research of the Spanish Ministry of Science and Innovation under grant PID2022-136828NBC41/AEI/10.13039/501100011033/, and by ‘ERDF A way of making
Europe’, by the European Union, and through the ‘Unit of Excellence
Maria de Maeztu 2020–2023’ award to the Institute of Cosmos
Sciences (CEX2019-000918-M). 
\end{acknowledgements}



\bibliographystyle{aa}
\bibliography{main}

\end{document}